# Evolution of electronic structure in pristine and hole-doped kagome metal RbV$_3$Sb$_5$

Jiawei Yu[†,‡], Kebin Xiao[†,‡], Yonghao Yuan[†,‡], Qiangwei Yin[§], Zhiqiang Hu[†,‡], Chunsheng Gong[§], Yunkai Guo[†,‡], Zhijun Tu[§], Hechang Lei[§,\*], Qi-Kun Xue[†,‡,⊥,∥,\*] and Wei Li[†,‡,\*]

[†]*State Key Laboratory of Low-Dimensional Quantum Physics, Department of Physics, Tsinghua University, Beijing 100084, China*

[‡]*Frontier Science Center for Quantum Information, Beijing 100084, China*

[§]*Department of Physics and Beijing Key Laboratory of Opto-electronic Functional Materials & Micro-nano Devices, Renmin University of China, Beijing 100872, China*

[⊥]*Beijing Academy of Quantum Information Sciences, Beijing 100193, China*

[∥]*Southern University of Science and Technology, Shenzhen 518055, China*

**Corresponding authors**

Wei Li Email: weili83@tsinghua.edu.cn

Hechang Lei Email: hlei@ruc.edu.cn

Qi-Kun Xue Email: qkxue@mail.tsinghua.edu.cn

**ABSTRACT:**

We report on *in situ* low-temperature (4 K) scanning tunneling microscope measurements of atomic and electronic structures of the cleaved surfaces of an alkali-based kagome metal RbV$_3$Sb$_5$ single crystals. We find that the dominant pristine surface exhibits Rb-1×1 structure, in which a unique unidirectional √3$a_0$ charge order is discovered. As the sample temperature slightly rises, Rb-√3×1 and Rb-√3×√3 reconstructions form due to desorption of surface Rb atoms. Our conductance mapping results demonstrate that Rb desorption not only gives rise to hole doping, but also renormalizes the electronic band structures. Surprisingly, we find a ubiquitous gap opening near the Fermi level in tunneling spectra on all the surfaces despite their large differences of hole-carrier concentration, indicating an orbital-selective band reconstruction in RbV$_3$Sb$_5$.

Electronic correlation and topology are pivotal problems in recent studies of condensed matter physics. Kagome system [1-15] provides an ideal platform for understanding these quantum phenomena, thanks to its unique lattice structure, in which electrons not only experience lattice frustration but also accumulate non-trivial Berry phase by hopping. Recently, a new family of kagome metals AV$_3$Sb$_5$ (A = K, Rb, Cs) has been discovered [16,17], which shows superconductivity with the highest transition temperature ($T_c$) of 2.5 K in CsV$_3$Sb$_5$. Several exotic quantum phenomena have been discovered in AV$_3$Sb$_5$ owing to the interplay between charge order [18-20], $Z_2$ topology [17,21,22], superconductivity [16,17,23-25] and the symmetry breaking of nematicity [19,26-28]. For example, the 2 × 2 charge order is considered to be a chiral flux phase with the time-reversal symmetry breaking [18,19,29], which may be responsible for the observed anomalous Hall effect [30,31] and topological edge state [32]. Evidence of topological surface states are reported in quantum transport [33] and angle-resolved photoemission spectroscopy (ARPES) measurements [34]. With the assistance of charge order, the topological surface states may cooperate with superconductivity [34], resulting in topological and anomalous superconductivity [22,23,25,35-38]. These features have generated considerable excitement as well as puzzles, such as the pairing symmetry of superconductivity, the driving force of charge order and the newly developed symmetry breaking phases.

Investigation of the electronic band structure is essential to reveal the underlying mechanisms of those quantum phenomena. Several ARPES groups measured the band structure of AV$_3$Sb$_5$ [34,39-42]. However, there exists a large deviation in the reported band evolution with temperature. One possible reason is that the alkali-metal atoms are easy to migrate and diffuse on the surface at higher temperature, which not only affects the chemical potential of the material, but also induces structural and electronic reconstructions [34,39,42]. The hole carriers, generated by the desorption of alkali-metal atoms, can dramatically affect the superconductivity [43]. It also moderately suppress the charge order in CsV$_3$Sb$_5$ by reduce transition temperature from 95 K to 74 K via orbital-selective hybridization to the related bands [43]. Given the alkali-metal-induced complexity of the surface morphology, exploration of the intrinsic atomic and electronic structure of the pristine cleaved surface of $AV_3Sb_5$ is highly desired.

$RbV_3Sb_5$ is a layered kagome superconductor [24] ($T_c$ ~ 0.9 K) with space group $P6/mmm$ and hexagonal lattice constants $a_0 = 5.47$ Å and $c_0 = 9.07$ Å. Its crystal structure is schematically shown in Figure 1a. The interlayer kagome $V_3Sb$ lattices are sandwiched by two layers of Sb honeycomb and two outer layers of Rb hexagonal lattice, respectively. To avoid the diffusion of alkali-metal atoms and obtain the pristine surface, we cleaved $RbV_3Sb_5$ single crystal at 4 K and then performed *in situ* scanning tunneling microscopy (STM) measurement on the fresh surface. The dominant pristine surface after the sample cleavage is the Rb-1×1 plane (Figure 1b-e), on which standing waves induced by impurities are clearly observed (Figure 1b). The STM images with atomic resolution (Figure 1c, d) exhibit 1 × 1 hexagonal lattice structure (indicated by white spheres in Figure 1a, c-e), upon which the widely studied 2 × 2 charge order [18-20, 29, 35, 44] are superposed. A unique unidirectional $\sqrt{3}a_0$ charge order (Figure 1e) is observed, indicating the further symmetry breaking of the Rb plane. The Sb terminated surface is also discovered, which shows honeycomb lattice with 2 × 2 charge order and unidirectional $4a_0$ supermodulation (see Figure 1f, g). The 2 × 2 charge order widely exists in the fully Rb-covered and bared Sb surfaces, indicating its robustness against surface doping at low temperature. Both of the pristine surfaces of $RbV_3Sb_5$ show unidirectional superstructures, corresponding to the observed nematicity [19, 26-28].

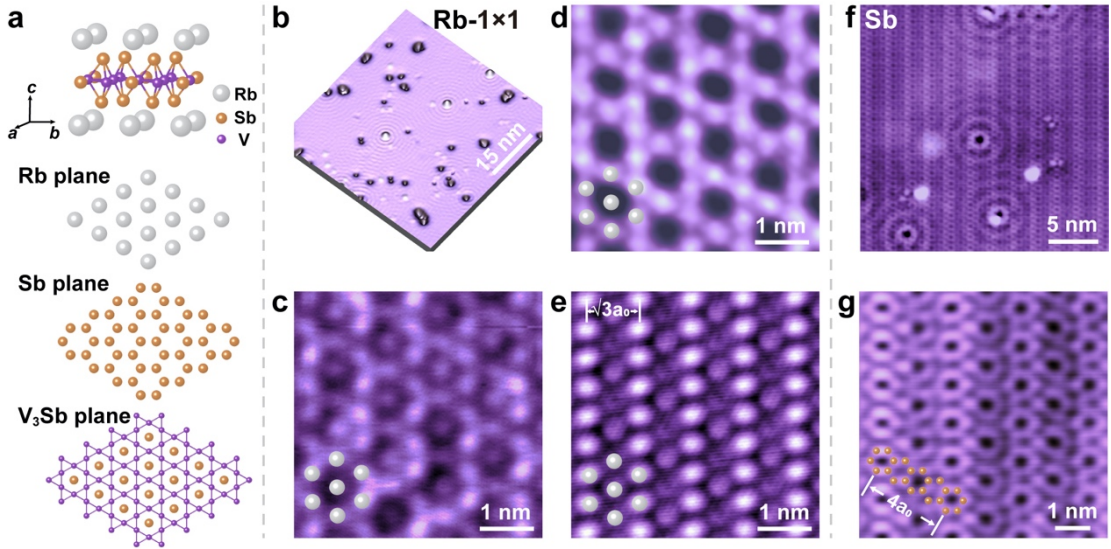

**Figure 1.** Pristine cleaved surfaces of $RbV_3Sb_5$. (a) Crystal structure of $RbV_3Sb_5$. b-c, STM topographic images of pristine cleaved Rb-terminated surface. Standing waves are visible in (b) (50 nm × 50 nm, bias voltage $V_s = 20$ mV, tunnelling current $I_t = 0.5$ nA). (c-e) Atomic resolution of the surface obtained under different STM setup conditions. Rb-1×1 hexagonal lattice is shown in (c) (4.5 nm × 4.5 nm, $V_s = -0.5$ mV, $I_t = 1.0$ nA) and (e) (4.5 nm × 4.5 nm, $V_s = -0.2$ mV, $I_t = 1.0$ nA). White spheres denote the Rb hexagonal lattice. 2 × 2 charge order is clearly shown in (d) (4.5 nm × 4.5 nm, $V_s = 150$ mV, $I_t = 0.8$ nA). The charge order is also superposed on the Rb-1×1 lattice in (c) and (e). A unidirectional superstructure with the period of $\sqrt{3}a_0$ is observed in (e). (f,g) STM

topographic images of pristine cleaved Sb-terminated surface. Standing waves are visible in f (25 nm × 25 nm, $V_s$ = 50 mV, $I_t$ = 0.8 nA). Sb honeycomb is shown in (g) (7 nm × 7 nm, $V_s$ = 10 mV, $I_t$ = 0.3 nA), which are denoted by orange spheres. A $4a_0$ unidirectional superstructure is observed in the Sb-terminated surface as well.

Considering the symmetry of the RbV$_3$Sb$_5$ lattice structure (Figure 1a), one may expect that the easily cleaved surface should be the Rb plane between two adjacent Sb honeycomb layers, and half amount of Rb atoms could be preserved on each side of the two cleaved surfaces. This is in sharp contrast to our observation, the full coverage of Rb atoms (forming the 1 × 1 lattice) on the freshly cleaved surface. This finding demonstrates that an additional symmetry breaking indeed occurs along the *c*-axis, consistent with the development of three-dimensional charge ordering [22, 45].

To study the temperature-dependent reconstruction of topmost Rb layer, we cleaved the samples at 77 K and at room temperature, respectively. As the cleavage temperature rises, the Rb coverage decreases due to the desorption of Rb atoms, leading to the formation of Rb-√3×1 (half coverage of Rb-1×1 plane) and Rb-√3×√3 (coverage ~ 1/3) reconstructions (Figure 2a). For the coverage of 1/2, which is commonly obtained for the cleavage at 77 K, a 2 × 2 superstructure is visible at negative bias (Figure 2b, c). At positive bias, individual Rb atoms appear and form √3 × 1 reconstruction (Figure 2d), in which the coverage ~ 1/2 can be immediately calculated (see the schematic in Figure 2a as well). For the cleavage at room temperature, large area with Rb-√3×√3 reconstruction comes out (Figure 2e). Here, the 2 × 2 charge order doesn't present due to its incompatibility to √3×√3 reconstruction. But it should still exist in the underneath Sb layer.

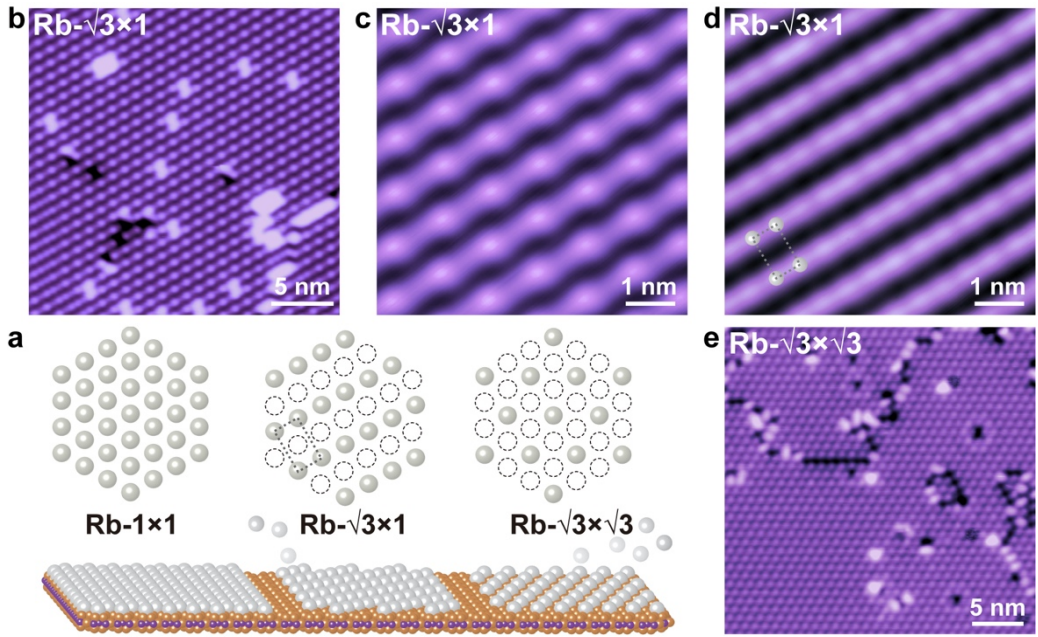

**Figure 2.** Reconstructions of Rb-terminated surface due to desorption of Rb atoms. (a) Schematic of three types of surface structures of Rb-terminated surface. (b-d) Atomically resolved STM

topographic images of Rb-√3×1. (b) and (c) show a 2 × 2 superstructure. Rectangular √3×1 reconstruction is revealed in (d). (e) Atomically resolved STM topographic images of Rb-√3×√3, which has lower coverage of Rb on the surface. STM setup: (b) 25 nm × 25 nm, bias voltage $V_s$ = -100 mV, tunnelling current $I_t$ = 0.2 nA; (c) 6.3 nm × 6.3 nm, $V_s$ = -300 mV, $I_t$ = 0.8 nA; (d) 6.3 nm × 6.3 nm, $V_s$ = 20 mV, $I_t$ = 0.8 nA; (e) 30 nm × 30 nm, $V_s$ = -500 mV, $I_t$ = 0.1 nA.

The temperature evolution of the surface reconstruction is consistent with annealing experiment of the sample (Figure 3). We transferred the sample out of STM head (4 K), and annealed it at room temperature; Then the sample was transferred back to STM and measured again. This is an annealing cycle (Figure 3a). Annealing process (with multiple cycles) is shown in Figure 3b. The pristine surface is Rb-1×1 with few Rb-vacancy (Figure 3c). With increased annealing time, a large amount of Rb-vacancies appears firstly (Figure 3d), then short-range of Rb-√3×1 surface starts to form with the continuous loss of Rb (Figure 3e, f), and long-range Rb-√3×√3 surface is obtained finally (Figure 3g). The temperature-dependent Rb reconstructions may give rise to various electronic structures.

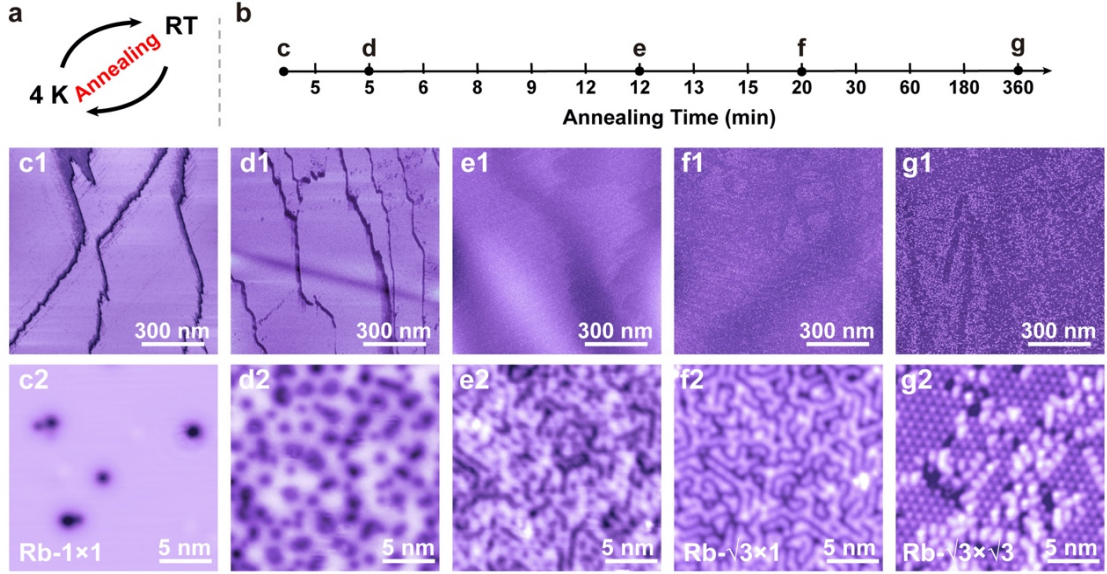

**Figure 3.** Annealing experiment of RbV$_3$Sb$_5$. (a) Schematic of an annealing cycle. (b) Annealing process of the sample. The number on time axis represents the annealing time for a single cycle. Representative STM topographies of the annealing experiment are shown in (d-g). (c1, c2) STM topographies of the pristine Rb surface. Set point: (c1) 1 um × 1 um, bias voltage $V_s$ = -300 mV, tunnelling current $I_t$ = 20 pA; (c2) 20 nm × 20 nm, $V_s$ = -200 mV, $I_t$ = 0.1 nA. (d1-g1) Topographies of Rb surface (1 um × 1 um, $V_s$ = -300 mV, $I_t$ = 20 pA) under the annealing cycle **d-g** labeled in (b). (d2-g2) Corresponding zoom-in topographic images of (d1-g1) ($V_s$ = -300 mV, $I_t$ = 20 pA), in which the desorption of Rb can be clearly observed: More Rb-vacancies appear in (d2); Short-range Rb-√3×1 reconstruction forms in (e2) and (f2); And finally Rb-√3×√3 reconstruction shows up in (g2).

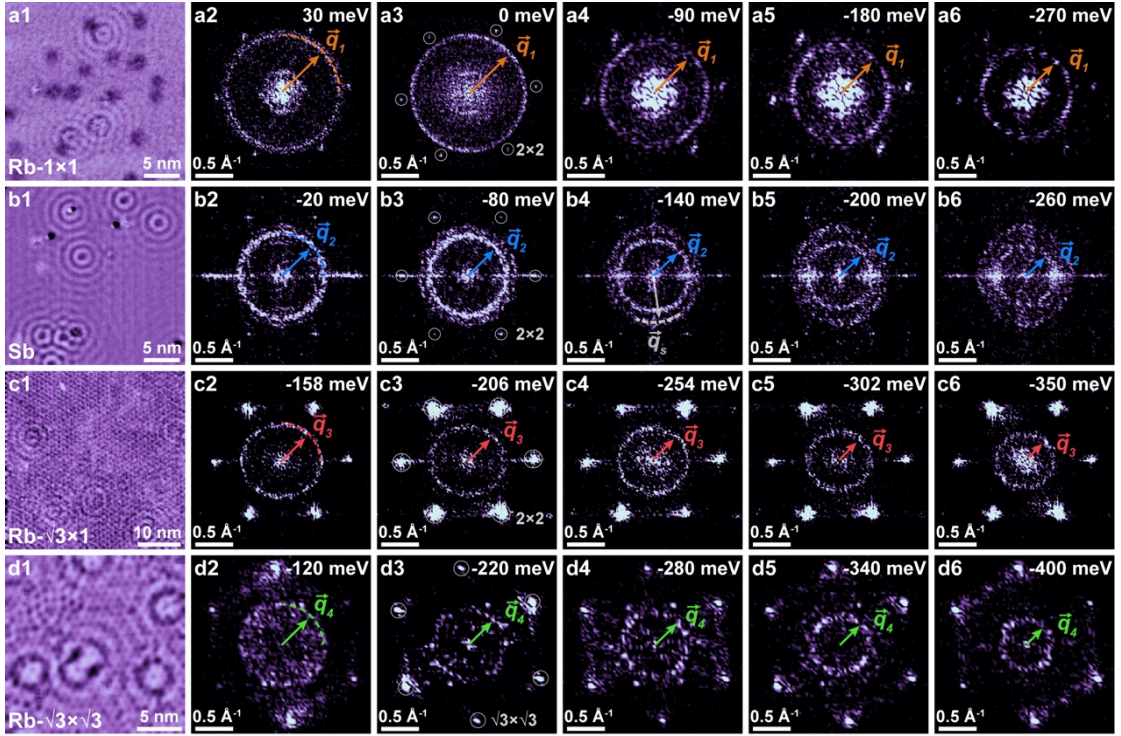

**Figure 4.** Quasiparticle interference results of four different surfaces. (a1) d$I$/d$V$ mapping of pristine Rb-1×1 surface at energy of -75 meV (25 nm × 25 nm). (a2-a6) FFT images of d$I$/d$V$ mapping of pristine Rb-1×1 surface at different energies. The circle-shaped structures indicate a spatially isotropic scattering and the scattering wave vector is marked by orange arrow. Set point: (a2, a3) bias voltage $V_s$ = -30 mV, tunnelling current $I_t$ = 0.5 nA; (a1, a4-a6) $V_s$ = 150 mV, $I_t$ = 0.4 nA. (b1) d$I$/d$V$ mapping of pristine Sb surface at energy of -60 meV (30 nm × 30 nm). (b2-b6) FFT images of d$I$/d$V$ mapping of pristine Sb surface at different energies. The spatially isotropic scattering wave vector is marked by blue arrow. Set point: (b1-b6) $V_s$ = 100 mV, $I_t$ = 1.0 nA. (c1) d$I$/d$V$ mapping of Rb-√3×1 surface at energy of -158 meV (40 nm × 40 nm). (c2-c6) FFT images of d$I$/d$V$ mapping of Rb-√3×1 surface at different energies, in which the scattering wave vector is marked by red arrow. Set point: (c1-c6) $V_s$ = -350 mV, $I_t$ = 0.8 nA. (d1) d$I$/d$V$ mapping of Rb-√3×√3 surface at energy of -180 meV (20 nm × 20 nm). (d2-d6) FFT images of d$I$/d$V$ mapping of Rb-√3×√3 surface at different energies, in which the scattering wave vector is marked by green arrow. Set point: (d1-d6) $V_s$ = -500 mV, $I_t$ = 1.0 nA.

We now turn to investigate the electronic structures of the above four surfaces with different termination. Scanning tunneling spectroscopy (STS or d$I$/d$V$ spectrum) probes the local electronic density of states (DOS), and d$I$/d$V$ mapping can show the spatial distribution of DOS at specific energy. Figure 4a1 shows a DOS map of Rb-1×1 surface at -75 meV, in which impurity-induced quasiparticle interference patterns are observed. The fast Fourier transform (FFT) images of DOS maps present a circle-shaped structure (Figure 4a2-a6), indicating a spatially isotropic scattering with a wave vector $\vec{q}_1$. The circle shrinks at higher energy below the Fermi level ($E_F$), demonstrating that the observed interference patterns originate from intra-pocket scattering of an electron-like band at the Γ point [39]. The signals of 2 × 2 charge order are also visible and marked by white circles in the FFT images. Quasiparticle interference shows similar behavior on Sb, Rb-√3×1 and Rb-√3×√3 surfaces (Figure 4b1-b6, c1-c6, d1-d6), while the magnitude of their corresponding scattering wave vector $\vec{q}_2$, $\vec{q}_3$ and $\vec{q}_4$ is various. On the Sb plane, an energy-independent isotropic scattering wave vector $\vec{q}_s$ larger than $\vec{q}_2$, is also observed (denoted by the grey arrow in Figure 4b4), and its origin is still unclear.

Examining the energy dispersion of each scattering wave vector $\vec{q}$ (Figure 5a) can provide rich information of the electronic structure, such as the Fermi momentum ($k_F$) and electron effective mass ($m^*$). As we mentioned before, the $\vec{q}$ here originates from the intra-pocket scattering of the electron-like band at the Γ point. Therefore, the relationship between the momentum ($\vec{k}$) and the $\vec{q}$ is $\vec{q} = 2\vec{k}$. Accordingly, the band bottoms of the corresponding electron-like pocket are determined by parabolic fitting of the $\vec{k}$ of each surface (see the fitting curves with different colors in Figure 5a). The band bottom of the Sb plane is the shallowest and located at -493 meV, which moves downward to -527 meV for Rb-√3×√3 surface, -551 meV for Rb-√3×1 surface and -666 meV for Rb-1×1 pristine plane, respectively. The tendency is consistent with the picture of Rb deficiency induced hole doping (see the red curve in the inset of Figure 5a).

In contrast, a sudden electron effective mass as well as the Fermi momentum jump occurs between the pristine Rb-1×1 and other reconstructed Rb surfaces (see the blue and green curves in the inset image of Figure 5a). The increase of effective mass indicates enhanced interactions between the electron-like band and other bands in RbV$_3$Sb$_5$ and the band renormalization of the pristine Rb-1×1 plane is indeed significantly different from other surfaces.

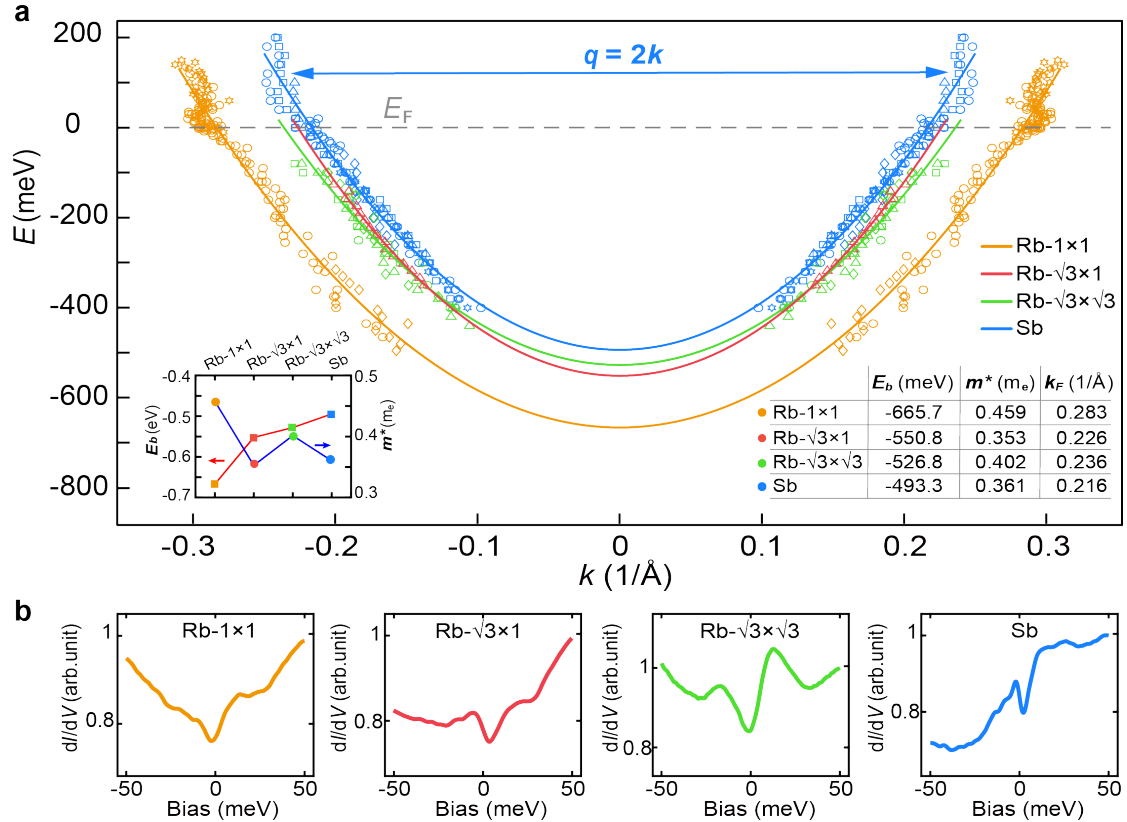

**Figure 5.** Electronic structures of four different surfaces. (a) Calculated electron-like bands at the Γ point of the four different surfaces. The data points of Rb-1×1, Rb-√3×1, Rb-√3×√3 and Sb surfaces are plotted in orange, red, green and blue, respectively. Different shapes of the data points correspond to different samples or locations. The details of the calculation are described in the main text. The values of energy of band bottom ($E_b$), electron effective mass ($m^*$) and Fermi momentum ($k_F$) of the electron-like band of the four surfaces are summarized in the inset table. The inset image shows the evolution of $E_b$ (red curve), $m^*$ (blue curve) and $k_F$ (green curve). (b) d$I$/d$V$ spectra of the four surfaces. A V-shaped gap-like feature is observed in all the spectra. Set point: Rb-1×1, bias voltage $V_s$ = 50 mV, tunnelling current $I_t$ = 0.4 nA; Rb-√3×1, $V_s$ = 50 mV, $I_t$ = 0.2 nA; Rb-√3×√3, $V_s$ = 50 mV, $I_t$ = 0.2 nA; Sb, $V_s$ = 50 mV, $I_t$ = 0.5 nA.

Surprisingly, STS shows similar V-shaped gap feature near $E_F$ on all of the four surfaces (Figure 5b), despite the large differences of chemical potential according to our DOS mapping results. The gap opening here is attributed to the 2 × 2 charge order, which is considered to be closely related to the electronic states near the M point [40, 46]. Since the electronic bands around the Γ and M points have different orbital component [39], the insensitivity of the gap-like feature to carrier doping indicates an orbital selective doping effect in RbV$_3$Sb$_5$. Rather than hybridizing with the $d$-orbital bands of V atoms around the M point, the hole carriers from Rb vacancies, in close proximity to the Sb layer, mainly affect the electron-like band at the Γ point, which is dominated by Sb $p_z$-orbital. Our finding here is consistent with the observation of orbital-selective band reconstruction in ARPES measurement [42].

We systematically investigate the evolution of the structural and electronic reconstruction of Rb plane with the loss of Rb atoms in RbV$_3$Sb$_5$ single crystal. We unambiguously demonstrate that the Rb deficiency not only provides hole carries to the sample, but also renormalizes the electronic structures. It selectively dopes the antimony $p_z$-band near Γ point and barely influences the vanadium $d$-orbital bands at M point. Since the desorption and reconstruction of Rb atoms are closely related to temperature, such complexity needs to be considered for the electronic structure measurements. We believe our discovery provides a reasonable explanation for the large deviation of band evolutions obtained from different ARPES groups [34, 39-42]. Meanwhile, the robust bands near M point gives rise to a ubiquitous 2 × 2 charge order and gap opening at low temperature. This charge order is reported to assist the cooperation between topological surface state and superconductivity [34]. Since superconductivity usually responses to surface doping [43], we expect that the amount of alkali-metal might serve as a tuning knob to separately control topological superconductivity from the impact of charge order. Therefore, identifying the pristine cleaved surface and investigating its electronic structure provide a solid starting point to understand and manipulate the complex physics in AV$_3$Sb$_5$.

**Methods.** RbV$_3$Sb$_5$ single crystals were grown and characterized as described in details in Ref. [24]. For 4 K cleavage, samples were transferred to the STM scanner (Unisoku). The samples were cleaved as they were cooled down to 4.2 K. A polycrystalline PtIr STM tip was used and calibrated using Ag island before STM experiments. STS data were taken by standard lock-in method. The feedback loop is disrupted during data acquisition and the frequency of oscillation signal is 811.0 Hz.

**Data availability.** The data that support the findings of this study are available from the corresponding author upon reasonable request.

## Author Contributions

W.L. and Q-K.X. conceived and supervised the research project. J.Y. and Y.Y. performed the STM experiments. Q.Y., C.G., Z.T. and H.L. grew the samples. W.L., J.Y., K.X., Y.Y., Z.H. and Y.G. analyzed the data. W.L. wrote the manuscript with input from all other authors.

## Notes

The authors declare no competing interests.

## ACKNOWLEDGEMENTS

We thank J. He, Y. Zhang, Z. Wang and J. Hu for inspiring discussions. The experiments were supported by the National Science Foundation (No. 51788104, No. 11674191), Ministry of Science and Technology of China (No. 2016YFA0301002) and the Beijing Advanced Innovation Center for Future Chip (ICFC). W. Li was supported by Beijing Young Talents Plan and the National Thousand-Young-Talents Program. H. Lei was supported by Ministry of Science and Technology of China (Grant No. 2018YFE0202600 and 2016YFA0300504), National Natural Science Foundation of China (Grant No. 11822412 and 11774423), Beijing Natural Science Foundation (Grant No. Z200005).